\documentclass[graybox]{svmult}

\usepackage{mathptmx}       
\usepackage{helvet}         
\usepackage{courier}        
\usepackage{type1cm}        
\usepackage{makeidx}         
\usepackage{graphicx}        
\usepackage{multicol}        
\usepackage[bottom]{footmisc}

\makeindex             

\begin{document}

\title*{Current state of the modelling of photospheric activity}
\author{A.~F.~Lanza}
\institute{ \at INAF-Osservatorio Astrofisico di Catania, via S.~Sofia, 78, 95125 Catania, Italy, \\  \email{nuccio.lanza@oact.inaf.it} }
%
%
\maketitle

\abstract*{
I briefly review the current state of the modelling of photospheric activity based on the high-precision optical light curves obtained with MOST, CoRoT, and Kepler. These models can be used to search for active longitudes where activity is preferentially concentrated, estimate the amplitude of  stellar differential rotation, and look for short-term activity cycles as, e.g., in the case of CoRoT-2. In the case of a late-type star accompanied by a transiting hot Jupiter, the small light modulations observed during transits when a dark spot is occulted by the disc of the planet are also briefly considered. They can be used to derive information on individual active regions as well as on  stellar rotation and the  spin-orbit alignment of the system.
}

\abstract{
I briefly review the current state of the modelling of photospheric activity based on the high-precision optical light curves obtained with MOST, CoRoT, and Kepler. These models can be used to search for active longitudes where activity is preferentially concentrated, estimate the amplitude of stellar differential rotation, and look for short-term activity cycles as, e.g., in the case of CoRoT-2. In the case of a late-type star accompanied by a transiting hot Jupiter, the small light modulations observed during transits when a dark spot is occulted by the disc of the planet are also briefly considered. They can be used to derive information on individual active regions as well as on  stellar rotation and the  spin-orbit alignment of the system.
}

\section{Introduction}
\label{sec:1}

In the Sun, we can study the photosphere in detail and map its brightness inhomogeneities associated with convective motions, oscillations, and magnetic fields. In particular, magnetic fields give rise to solar \emph{active regions} characterized by dark spots, bright faculae and an enhanced network. 

The transit of an active region across the solar disc as our star rotates produces a characteristic modulation of the optical flux with a rise when the region appears close to the approaching limb, a dip when it is closest to the central meridian, and another rise when it reaches the receding limb. The flux rises are associated with faculae that are brighter close to the limb and do not show an appreciable contrast when they are close to the disc centre, while the flux dip is produced by the dark sunspots whose photometric effect is maximum when their projected area is maximum, i.e., when the active region is closest to the centre  of the disc (cf., e.g., \cite{FrohlichLean04}).

The collective behaviour of solar active regions  shows a remarkable regularity. The total sunspot area oscillates with a mean period of $\sim 11$~yr that is the main feature of the solar \emph{activity cycle}.  The latitudes where sunspot groups form migrate along the cycle, with the first groups appearing at $\sim 30^{\circ}-35^{\circ}$ and then migrating towards the equator as the cycle progresses. At the maximum of activity, sunspots form at a latitude of $\sim 15^{\circ}-20^{\circ}$ to reach a minimum latitude of $5^{\circ}-10^{\circ}$ at the end of the cycle, when the sunspot groups associated with the next cycle begin to appear at high latitudes. 

The variation of the sunspot area is accompanied by a variation of the eigenfrequencies of the solar p-mode oscillations and of the amplitude and full width at half maximum (hereafter FWHM) of their peaks in the power spectrum. Since we wish to compare the solar behaviour with that of distant stars, we focus on  low-degree modes ($\ell =0, 1,2$) that can be detected in disc-integrated observations. The variation of their eigenfrequencies along the previous solar cycle 23 and the onset of the present solar cycle 24 has been reported  by \cite{Salabertetal09}, while \cite{Jimenez-Reyesetal03} have presented the variation of the amplitude and FWHM of the mode peaks in the power spectrum. Note that the onset of  the new cycle 24 has been  apparent in the $\ell = 0$ and $2$ modes, while it has not been detected in the $\ell=1$ modes. This is related to the spatial distribution of the magnetic perturbations in the Sun because the $\ell=0, 2$ modes are more sensitive to the high-latitude perturbations, while the $\ell=1$ modes respond mainly to the low-latitude perturbations. Since the magnetic fields are localized at high latitudes at the beginning of a new cycle, their effect can be detected in the $\ell=0, 2$ modes even before the first sunspot groups appear (cf. \cite{Salabertetal09} for details). 

Analogous phenomena are expected in  late-type stars with a level of magnetic activity  comparable with  the Sun. They have outer convection zones that host hydromagnetic dynamos generally leading to cycles \cite{Baliunasetal95} and display convection-driven  p-mode oscillations. Indeed, \cite{Garciaetal10} have detected a dependence of the eigenfrequencies and  the mode amplitudes on the level of magnetic activity in the F-type star HD~49933, recently observed by the CoRoT (Convection, Rotation and Transits) space telescope. 

Taking the Sun as a template, to perform a comparative study of the impact of activity on p-mode oscillations in  distant stars, we need to derive: a) the variation of the total spotted area vs. time, to trace stellar activity cycles; b) the latitude of the active belts vs. the phase of the activity cycle, assuming that there is a regular migration as in the Sun (see \cite{Mossetal11} for a detailed discussion of the possible behaviours). In the next sections, we shall introduce the methods to map  stellar photospheres that allow us to achieve these goals and present some application to the case of CoRoT targets. The advantages of stars accompanied by transiting hot Jupiters will be also emphasized.  

\section{Methods to map the photospheres of late-type stars}

To map stellar photospheres, we can apply methods based on spectroscopic and/or photometric data. Doppler Imaging (hereafter DI) is the most powerful technique based on spectroscopy, while  photometric spot modelling is based on  fitting  wide-band light curves. 

DI is based on the line profile distortions induced by surface brightness inhomogeneities that can be resolved when the stellar rotation is the main source of line broadening \cite{VogtPenrod83}. In practice, this implies  $v \sin i \geq 10-15$~km~s$^{-1}$ in a main-sequence star. In G or K-type dwarf stars, this corresponds to a maximum rotation period of $ \sim 4-5$~days that implies an optical variability induced by magnetic activity of at least $0.02-0.05$~mag. This may hamper the detection of p-mode oscillations as recently found by \cite{Chaplinetal11} by analysing a sample of Kepler targets (see also \cite{Mosseretal09a} for the case of the active CoRoT target HD~175726). Another limitation of  DI is the lack of sensitivity to a spot distribution  that is uniform on length scales comparable with or larger than its spatial resolution. This poses a severe limitation on the measurement of the total spotted area in stars that are slowly rotating, as most of the asteroseismic late-type targets. Nevertheless, DI is capable of retrieving the latitude of starspots which is not possible with  spot modelling, except in very favourable circumstances, and can be applied to spectropolarimetric observations of magnetic sensitive lines to map the surface magnetic fields and look for magnetic cycles (Zeeman Doppler Imaging, see, e.g., \cite{Faresetal09}).
	 
Spot modelling techniques are based on the reproduction of the rotational modulation of the optical flux induced by  starspots. If there is a single spot on a  rotating star, the time of light minimum can be used to derive its longitude and the maximum deficit of the flux  gives a measure of the projected area of the spot. If the star is not viewed equator-on, the duration of the spot transit across the stellar disc can be used to estimate its latitude. As a matter of fact, several spots are usually present on the surface of an active star which makes the light curve inversion non-unique and often unstable because small changes in the data imply large changes in the derived spot positions and unprojected areas. In principle, a light curve is a one-dimensional data set, thus the derivation of a two-dimensional map of the photosphere without some additional assumptions is impossible. 
Nevertheless, the longitude of the spots and their total projected area are robust quantities that can usually be retrieved  unambiguously from a high-precision  light curve. This has been demonstrated in the case of the Sun, for which we have spatially resolved maps of the photosphere,  by e.g., \cite {Lanzaetal07}.

To alleviate the non-uniqueness and  instability problems, two approaches have been applied. The first adopts a limited number of spots, usually up to 3 or 4, and fits a limited number of free parameters including the spot coordinates and areas  \emph{(few-spot modelling)}.  Several  works have discussed the  technique beginning from the earliest in the '70 and '80 \cite{TorresFerraz-Mello73,Rodonoetal86}, to the first application to time series of data covering several consecutive rotations \cite{StrassmeierBopp92}, and finally to the time-series photometry obtained by space-borne telescopes like MOST (the Microvariability and Oscillation of Star satellite, \cite{Rucinskietal03}). 

The other approach to  spot modelling consists in assuming a continuous distribution of small spots,  parameterized by a filling factor $f$ giving the fraction of the surface occupied by the spots at each given location. To avoid the instability and non-uniqueness of the solution, the method needs  a priori assumptions on the distribution of $f$ to regularize the solution. The best approach, in the sense that it reproduces the distribution of sunspot groups in the Sun \cite{Lanzaetal07}, is based on the \emph{Maximum Entropy regularization} (hereafter ME), where we look for the map with the maximum Shannon entropy among the infinite distributions of the filling factor that fit the light curve down to the precision of the measurements. From ME maps, the distribution of the filling factor vs. longitude and the variation of the total spotted area vs. time are derived. The information on spot latitude is usually very limited or absent in a light curve (cf. e.g., \cite{Lanzaetal09a}), so we should not attempt to retrieve it from ME maps.

Finally, we should mention the possibility of using the occultations of the starspots by a transiting planet as a high-pass spatial filter to map the fine structure of the spot distribution within the occulted band. The method has been successfully applied to, e.g., CoRoT-2 by \cite{Silva-Valioetal10,Silva-ValioLanza11} and provides information  on the characteristics of individual spots as well as on stellar differential rotation and the spin-orbit alignment of the planetary system \cite{Nutzmanetal11}.

\section{Results}
\label{results}

Long-term photometry allows us to derive stellar butterfly diagrams by tracing the systematic variation of the stellar rotation period as derived from the optical flux modulation vs. the time.  The variation of the total spotted area can  be measured simultaneously by, e.g., the variation of the mean light level from one season to the next,  allowing us to trace the phase of the cycle and, in combination with the variation of the rotation period, the direction of migration of the spots \cite{MessinaGuinan03}. It is interesting to note that in some stars the activity belts may migrate polewards (see, e.g., \cite{Mossetal11}). In stars that rotate as slowly as the Sun, it is difficult to measure the rotation period using photospheric spots because they evolve on time scales of $10-15$ days in the Sun, i.e., significantly shorter than the rotation period itself. Chromospheric faculae live longer and are  better tracers to look for the systematic variations associated with the solar cycle (see \cite{HempelmannDonahue97} and references therein).

An estimate of the starspot temperatures based on multi-wavelength ground-based photometry has been presented by \cite{AmadoZboril02}. The high-precision space-borne photometry provided by MOST has allowed to measure the surface differential rotation of several late-type stars \cite{Crolletal06}.  Among the latest results, those obtained by \cite{Mosseretal09b} are particularly interesting from the point of view of asteroseismology because they refer to F-type main-sequence stars observed by CoRoT to look for p-mode oscillations. In addition to the spot lifetimes, they estimate the amplitude of the surface differential rotation and the inclination of the spin axis of the star to the line of sight with a typical uncertainty of $20^{\circ}-30^{\circ}$. Asteroseismology can do much better providing us with an inclination with an uncertainty of $5^{\circ}-10^{\circ}$ in the case of such stars that rotate $4-5$ times faster than the Sun \cite{GizonSolanki03}. In turn,  this can be used to improve the estimate of the differential rotation from the spot modelling. 

The activity of several planet-hosting stars discovered by CoRoT has been investigated. CoRoT-2 shows a short-term activity cycle with a period of about 29 days that could be analogous to the solar Rieger cycles \cite{Oliveretal98,Lou00,Zaqarashvilietal10} or may be induced by some kind of star-planet interaction \cite{Lanzaetal09a,Lanza11}.  Active longitudes have been detected and estimates of the surface differential rotation have been obtained for the other targets CoRoT-4 \cite{Lanzaetal09b}, CoRoT-7 \cite{Lanzaetal10}, and CoRoT-6 \cite{Lanzaetal11}. 

The power of starspot occultation during transits in active stars accompanied by hot Jupiters has  been shown by \cite{Sanchis-OjedaWinn11} in the case of HAT-P-11 as observed by the Kepler space telescope. In addition to a quite accurate estimate of the inclination of the stellar spin, they have obtained a measurement of the true misalignment of the system by combining spot modelling with  spectroscopic measurements of the Rossiter-McLaughlin effect. The latitude of the  starspots has been constrained thanks to the large misalignment of the system. This  offers a new opportunity to build stellar butterfly diagrams  by means of the long-term Kepler photometry.  

\section{Discussion and conclusions}

The recent detection of a dependence of the p-mode properties on the activity level in HD~49933,  similar to that observed in the Sun, has revived the interest in studying stellar activity in late-type stars with outer convection zones. Those stars display p-mode oscillations excited by convection as well as activity phenomena produced by  dynamo-generated magnetic fields. We have illustrated as the same high-precision photometric data collected to measure p-mode oscillations (and in several cases to search for planetary transits) can be used to model the distribution of  starspots vs. longitude in a stellar photosphere and  trace the variation of their area and migration along the activity cycle. This will allow us a better understanding of the impact of magnetic activity on p-mode oscillations in solar-like stars thanks to the long-term photometry provided by the CoRoT and Kepler missions (see  the talk by S. Mathur). Spot modelling can significantly benefit from the determination of the stellar inclination made possible by asteroseismology. This in turn improves  the estimation of the stellar differential rotation which is of fundamental importance to understand the rotational splitting of the p-mode frequencies and the operation of the stellar dynamo. Stars accompanied by a transiting planet offer new opportunities for the spot modelling thanks to the occultation of the spots by the planet and are also extremely interesting asteroseismic targets (see the talk by A. Moya).

\begin{acknowledgement}
The author wishes to thank the SOC of the 20th Stellar Pulsation Conference for their kind invitation to attend the Conference and their nice hospitality in Granada, and Prof.~B.~Mosser for interesting discussions. 
\end{acknowledgement}


\begin{thebibliography}{99.}%
%

\bibitem[Amado \& Zboril (2002)]{AmadoZboril02} 
Amado, P.~J., \& Zboril, M.\ 2002, A\&A, 381, 517 



\bibitem[Baliunas et al. (1995)]{Baliunasetal95} 
Baliunas, S.~L., Donahue, R.~A., Soon, W.~H., et al.\ 1995, ApJ, 438, 269 


\bibitem[Chaplin et al. (2011)]{Chaplinetal11} Chaplin, W.~J., Bedding, T.~R., Bonanno, A., et al.\ 2011, ApJL 732, L5 

\bibitem[Croll et al. (2006)]{Crolletal06} 
Croll, B., Walker,  G.~A.~H., Kuschnig, R., et al.\ 2006, ApJ, 648, 607 


\bibitem[Fares et al. (2009)]{Faresetal09} 
Fares, R., Donati, J.-F., Moutou, C., et al.\ 2009, MNRAS 398, 1383 

\bibitem[Fr{\"o}hlich \& Lean (2004)]{FrohlichLean04}
Fr{\"o}hlich, C.~\& Lean, J., 2004, Astronomy and Astrophysics Review, 12, 273. 

\bibitem[Garc{\'{\i}}a et al. (2010)]{Garciaetal10} 
Garc{\'{\i}}a, R.~A., Mathur, S., Salabert, D., et al.\ 2010, Science, 329, 1032 

\bibitem[Gizon \& Solanki (2003)]{GizonSolanki03} 
Gizon, L., \& Solanki, S.~K.\ 2003, ApJ 589, 1009 



\bibitem[Hempelmann  \& Donahue (1997)]{HempelmannDonahue97} 
Hempelmann, A., \& Donahue, R.~A.\ 1997, A\&A, 322, 835 



\bibitem[Jim{\'e}nez-Reyes et al. (2003)]{Jimenez-Reyesetal03}  Jim{\'e}nez-Reyes, S.~J., Garc{\'{\i}}a, R.~A., Jim{\'e}nez, A., \& Chaplin, W.~J.\ 2003, ApJ, 595, 446 

\bibitem[Lanza (2011)]{Lanza11} 
Lanza, A.~F.\ 2011, Ap\&SS, 336, 303

\bibitem[Lanza et al. (2007)]{Lanzaetal07} 
Lanza, A.~F., Bonomo, A.~S., \& Rodon{\`o}, M.\ 2007, A\&A, 464, 741 


\bibitem[Lanza et al. (2009a)]{Lanzaetal09a} 
Lanza, A.~F., Pagano, I., Leto, G., et al.\ 2009a, A\&A 493, 193 

\bibitem[Lanza et al. (2009b)]{Lanzaetal09b} 
Lanza, A.~F., Aigrain, S., Messina, S., et al.\ 2009b, A\&A, 506, 255 

\bibitem[Lanza et al. (2010)]{Lanzaetal10} 
Lanza, A.~F., Bonomo, A.~S., Moutou, C., et al.\ 2010, A\&A, 520, A53 

\bibitem[Lanza et  al. (2011)]{Lanzaetal11} 
Lanza, A.~F., Bonomo, A.~S., Pagano, I., et al.\ 2011, A\&A, 525, A14 

\bibitem[Lou (2000)]{Lou00} 
Lou, Y.-Q.\ 2000, ApJ, 540, 1102 


\bibitem[Messina \& Guinan (2003)]{MessinaGuinan03} 
Messina, S., \& Guinan, E.~F.\ 2003, A\&A, 409, 1017 

\bibitem[Moss et al. (2011)]{Mossetal11} 
Moss, D., Sokoloff, D., \& Lanza, A.~F.\ 2011, A\&A, 531, A43 

\bibitem[Mosser et al. (2009a)]{Mosseretal09a} 
Mosser, B., Michel, E., Appourchaux, T., et al.\ 2009a, A\&A 506, 33 


\bibitem[Mosser et al. (2009b)]{Mosseretal09b} 
Mosser, B., Baudin, F., Lanza, A.~F., et al.\ 2009b, A\&A 506, 245 


\bibitem[Nutzman et al. (2011)]{Nutzmanetal11} 
Nutzman, P.~A., Fabrycky, D.~C., \& Fortney, J.~J.\ 2011, ApJL, 740, L10 


\bibitem[Oliver et al. (1998)]{Oliveretal98} 
Oliver, R., Ballester,  J.~L., \& Baudin, F.\ 1998, Nature, 394, 552 

\bibitem[Rodon\'o et al. (1986)]{Rodonoetal86} 
Rodon\'o, M., Cutispoto, G., Pazzani, V., et al.\ 1986, A\&A 165, 135 


\bibitem[Rucinski et al. (2004)]{Rucinskietal03} Rucinski, S.~M., 
Walker, G.~A.~H., Matthews, J.~M., et al.\ 2004, PASP, 116, 1093 

\bibitem[Salabert et  al. (2009)]{Salabertetal09} 
Salabert, D., Garc{\'{\i}}a, R.~A., Pall{\'e}, P.~L., \& Jim{\'e}nez-Reyes, S.~J.\ 2009, A\&A, 504, L1 



\bibitem[Salabert et al. (2011)]{Salabertetal11} Salabert, D., R{\'e}gulo, C., Ballot, J., Garc{\'{\i}}a, R.~A., \& Mathur, S.\ 2011, A\&A, 530, A127 

\bibitem[Sanchis-Ojeda \& Winn (2011)]{Sanchis-OjedaWinn11} 
Sanchis-Ojeda, R., \& Winn, J.~N.\ 2011, ApJ,  743, 61



\bibitem[Silva-Valio \& Lanza (2011)]{Silva-ValioLanza11} 
Silva-Valio, A., \& Lanza, A.~F.\ 2011, A\&A, 529, A36 


\bibitem[Silva-Valio et al. (2010)]{Silva-Valioetal10} Silva-Valio, A., Lanza, A.~F., Alonso, R., \& Barge, P.\ 2010, A\&A 510, A25 



\bibitem[Strassmeier \& Bopp (1992)]{StrassmeierBopp92} 
Strassmeier, K.~G., \& Bopp, B.~W.\ 1992, A\&A, 259, 183 



\bibitem[Torres \& Ferraz Mello (1973)]{TorresFerraz-Mello73} 
Torres, C.~A.~O., \& Ferraz Mello, S.\ 1973, A\&A, 27, 231 

\bibitem[Vogt  \& Penrod (1983)]{VogtPenrod83} 
Vogt, S.~S., \& Penrod, G.~D.\ 1983, PASP, 95, 565 



\bibitem[Zaqarashvili et al. (2010)]{Zaqarashvilietal10} 
Zaqarashvili, T.~V., Carbonell, M., Oliver, R., \& Ballester, J.~L.\ 2010, ApJ, 709, 749 


\bigskip

\end{thebibliography}
\end{document}